\documentclass[debug,overfull,nocite]{epl}
\usepackage{amsmath}
\usepackage{amssymb}
\usepackage{graphicx}
\title{Suppressed absolute negative conductance and
generation of high-frequency radiation in semiconductor superlattices}
\shorttitle{suppressed ANC and generation}
\author{K. N. Alekseev\inst{1} \and
M. V. Gorkunov\inst{1,2} \and N. V. Demarina\inst{3}
\and T. Hyart\inst{1} \and N. V. Alexeeva\inst{1} \and A. V. Shorokhov\inst{1}
}
\institute{
\inst{1} Department of Physical Sciences, P.O. Box 3000, University of Oulu 90014, Finland\\
\inst{2} Institute of Crystallography, Russian Academy of Sciences, Moscow 119333, Russia\\
\inst{3} Nizhny Novgorod State University, 603950 Nizhny Novgorod, Russia\\
}
\begin{document}
\maketitle

\begin{abstract}
We show that space-charge instabilities (electric field domains)
in semiconductor superlattices are the attribute of absolute
negative conductance induced by small constant and large
alternating electric fields. We propose the efficient method for
suppression of this destructive phenomenon in order to obtain a
generation at microwave and THz frequencies in devices operating
at room temperature. We theoretically proved that an unbiased
superlattice with a moderate doping subjected to a microwave pump
field provides a strong gain at third, fifth, seventh, etc. harmonics of
the pump frequency in the conditions of suppressed domains.
\end{abstract}
\section{Introduction}
There exists a strong demand for miniature, solid-state, room
temperature operating sources and detectors of THz radiation
($0.3$-$10$ THz). The need is caused by a rapid progress of THz
sciences and technologies ranging from the astronomy to the
biosecurity \cite{thz-reviews}. Semiconductor superlattices (SLs)
\cite{esaki70}, operating in the miniband transport regime, are
interesting electronic devices demonstrating properties of both
nonlinear and active media. Nonlinearity of voltage-current (UI)
characteristic of SL gives rise to a generation of harmonics of
microwave and THz radiation \cite{esaki71,ignatov76}. On the other
hand, Bloch oscillations of electrons within a miniband of SL
cause an appearance of negative differential conductance (NDC) for
dc fields (voltages) larger than the critical Esaki-Tsu $E_c$
\cite{esaki70}. Remarkably, the static NDC should be accompanied
by a small-signal gain for ac fields of very broad frequency range
from zero up to several THz \cite{kss}. The feasible device,
employing such active media properties of dc-biased SL at room
temperature, is known as THz Bloch oscillator.
However, an existence of static NDC causes space-charge
instabilities and a formation of high-field domains inside SL
\cite{buettiker77}. These domains are believed to be destructive
for the Bloch gain. Recent top class experiment demonstrates only
decrease in THz absorption, but still not gain, in an array of
short dc-biased SLs \cite{savvidis}. Therefore, it is important to
consider modifications of this canonic scheme of generation in
order to suppress electric domains but preserve the high-frequency
gain in SLs. Recently, the experimental realizations of SL
oscillators with ac bias have been reported
\cite{renk-apl,renk-prl}.
\par
In the present Letter, we analyze theoretically the scheme of
high-frequency superlattice oscillator where instead of dc bias a
quasistatic ac electric field is used. Because for typical SLs the
characteristic scattering time at room temperature is of the order
of $100$ fs, the quasistatic interaction of ac field with miniband
electrons is well defined for the microwave frequencies. SL is
placed into the resonator tuned to the frequency of generation.
Here the ac field plays a twofold role: It suppresses the
space-charge instability inside SL and simultaneously pumps an
energy for generation and amplification at higher frequency.
\par
The main idea of our scheme for suppression of space-charge
instability is very simple. Suppose that the frequency of the pump
field $\omega$ is larger than the inverse characteristic time of
domain formation $\tau_{dom}^{-1}$. This requirement,
$\omega\tau_{dom}\gg 1$, imposes the lower limit for the values of
pump frequency $\omega$ in our scheme, which, however, still
easily belongs to the microwave range. Then, we show that the
electric instability in the ac-driven superlattice arises if the
slope of dependence of the time-averaged current on the applied dc
bias is negative (NDC for the time-average current). We note that
the only possibility to have NDC for the time-average current in
the limit of small dc bias is related to the existence of absolute
negative conductance (ANC), when the direction of dc current is
opposite to its direction in a normal conductor
(fig.~\ref{fig_anc}). ANC is known in various systems
\cite{nature},  but this nonequilibrium phenomenon requires very
special conditions, which as rule are quite difficult to satisfy.
In particular, only a few combinations of quasistatic ac fields
can induce ANC in SL. Thus, a high electric stability of the
oscillator's scheme is based on a simplicity of ANC state
suppression in unbiased or weakly dc-biased SLs.
\par
\begin{figure}
\onefigure[scale=0.3]{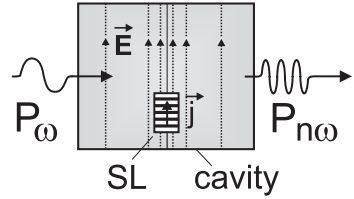}
\caption{\label{fig_device}Schematic representation of the
domainless oscillator: Generation at $n=3,7\ldots$.}
\end{figure}

We prove that within this stabilization scheme the only way to get
gain for both small- and large-signals, is to tune a resonator to
some particular harmonics of $\omega$ (fig.~\ref{fig_device}). The
physical processes in SL contributing the the net gain are the
parametric amplification of the field in resonator, seeded by the
the frequency multiplication of the pump field, as well as the
nonparametric absorption controlling the suppression of domains.

\section{Suppression of domains}
We consider a response of electrons, belonging to a single
miniband of SL, to the action of strong pump field
$E_{p}=E_{dc}+E_\omega\cos\omega t$, where dc bias $E_{dc}$ is
small and ac field is quasistatic $\omega\tau<1$ ($\tau$ is the
intraminiband relaxation time). This pump field creates the
voltage $U_{p}=E_{p}L$ across SL of the total length $L$.
We suppose also that due to a
nonlinear character of miniband transport, a signal
with amplitude $E_1<E_\omega$ and frequency $\omega_1>\omega$ is generated.
The total voltage across SL $U(t)$ is
\begin{equation}
\label{total_eq}
U=U_p(t)+U_s(t)=(U_{dc}+U_{\omega}\cos\omega t)+U_1\cos\omega_1 t .
\end{equation}
For $\omega\tau<1$, the dependence of the current through SL
$I(t)$ on the ac voltage $U(t)$ can be well described by the
Esaki-Tsu formula
\begin{equation}
\label{esaki-tsu_eq}
I=2 I_p(U/U_c)/[1+(U/U_c)^2],
\end{equation}
where $U_c=(\hbar L)/e a\tau$ is the critical voltage ($a$ is the SL period),
and $I_p\equiv I(U_c)$ is the peak current \cite{esaki70}.
\begin{figure}
\onefigure[scale=0.5]{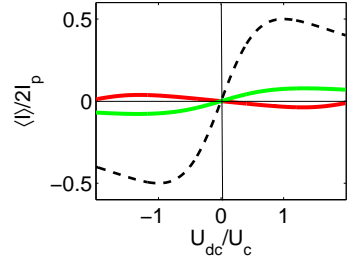}
\caption{\label{fig_anc}[color online] Dependence of time-averaged current $\langle I\rangle$
 on dc voltage $U_{dc}$ for the pump $U_{\omega}/U_c=9$ and for different strengths of signal
$U_1$ at frequency $7\omega$: Positive slope for
$U_1/U_c=0.9$ (green online), negative slope and ANC for $U_1/U_c=6.3$ (red online).
Esaki-Tsu characteristic ($U_{\omega}=0$) is shown as dashed line.}
\end{figure}
Now we turn to the derivation of electric stability conditions of SL device.
For $\omega\tau<1$ a space-time evolution of the electron density
$\rho(x,t)$, the current density $j(x,t)$ and the field $E(x,t)$ inside SL, driven by the given voltage
$U(t)$, can be well described by  the drift-diffusion model
\cite{bonilla-grahn}.
Set of equations consists of the current
equation $j=e \rho V(E)-D(E)\partial \rho/\partial x$, the Poisson
equation $\partial E/\partial x=4\pi\epsilon^{-1}(\rho-N)$ with $N$ being the doping density,
the relation to the applied voltage $U(t)=\int_0^L E(x,t)dx$, and the
continuity equation $e\partial\rho/\partial t+\partial j/\partial
x=0$ \cite{dd_review}. Dependence of the electron velocity $V$ on the local field
$E(x,t)$ is determined by the Esaki-Tsu formula as $V(E)=2 V_p(E/E_c)/[1+(E/E_c)^2]$ with
$E_c=\hbar/e a\tau$ and $V_p\equiv V(E_c)$. Following the Einstein relation the
diffusion coefficient is $D=k_B T V(E)/E$.
If the amplitude of $U(t)>U_c$, space-charge instabilities
periodically arise during the part of
period of ac voltage $T=2\pi/\omega$ when the SL is switched to the state with NDC.
\par
We suppose that condition $\omega\tau_{dom}\gg 1$ is satisfied.
This inequality guarantees that a charge accumulated during $T$ is always limited.
In this situation, fluctuations of $j$ and $\rho$ are small and we can use linear analysis
investigating stability of the system. The characteristic time $\tau_{dom}$ is of the order of
the dielectric relaxation time $\sim(\omega_{pl}^2\tau)^{-1}$
($\omega_{pl}=(4\pi e^2 N/ \epsilon
m_0)^{1/2}$ is the miniband plasma frequency, $N$ is the doping
density, $m_0$ is the electron mass at the bottom of miniband, $\epsilon$ is
the averaged dielectric constant of SL) \cite{dd_review}.
In more accurate approach $\tau_{dom}$ depends on the amplitude of ac field.
Our numerical simulations demonstrate that for typical semiconductor SL
with $N\simeq 10^{16}$ cm$^{-3}$ accumulated charge is small ($\omega\tau_{dom}\gtrsim 1$)
if $\omega/2\pi\gtrsim 100$ GHz. Decreasing doping density $N$ one can satisfy the condition using
lower pump frequencies.
\par
Linearizing the drift-diffusion model equations, we find that
small fluctuations of space-charge with a long wavelength will not
grow, if the dependence of the time-averaged current $\langle
I\rangle$ on the dc voltage $U_{dc}$ has positive slope, i.e.
\begin{equation}
\label{lambda_eq}
\Lambda=\langle I^\prime[U(t)]\rangle>0,
\end{equation}
where averaging $\langle\ldots\rangle$ is performed over period
$T$, prime means first taking the derivative in respect to
$U_{dc}$ and then assuming $U_{dc}\rightarrow 0$. Here and in what
follows, we will consider the current $I$ and the voltage
amplitudes $U_{\omega}$ in units of $I_{p}$ and $U_c$,
correspondingly. Combining eqs. (\ref{lambda_eq}) and
(\ref{esaki-tsu_eq}) for the quasistatic pump $U=U_p(t)$, we get
$\Lambda_p\equiv \Lambda(U_p)=(1+U_{\omega}^2)^{-3/2}$, which is
\textit{always positive} for all $U_{\omega}$. Therefore, SL
driven by a monochromatic and quasistatic ac field is stable
against small fluctuations of charge or field.
\par
We also considered the influence of signal fields on the stability
of the system calculating numerically the increment $\Lambda$ for
$U(t)$ given by eq.~(\ref{total_eq}) with $\omega_1=n\omega t$.
For $n=3$ we found that $\Lambda>0$ for all $U_1\leq
10$. Nevertheless, the UI characteristics with ANC, and as a
consequence with $\Lambda<0$, can exist for some particular values
of $U_\omega$ and $U_1$ in the cases $n=5$ and $n=7$
(fig.~\ref{fig_anc}).  However, for the practically important
range $U_1<U_1^{st}$ (see next section), we still found no ANC.
\par
We should distinguish our scheme of electric stabilization based
on avoidance of ANC in unbiased SLs from the free of domains
Limited Space Accumulation (LSA) regime, which is well-known in
the Gunn diodes (see \cite{copeland} for the theory and
\cite{kagan} for the theory and experiment) and can be applied to
SLs as well \cite{rieder}. This LSA regime works only for a large
dc bias and for a large ac field in the resonator. In contrast,
within our scheme it is possible to avoid a formation of domains
even for small-amplitude fields in resonator.
%
\section{Small-signal gain}
In the oscillator, SL is placed in a resonator providing a feedback only for the signal
with frequency $\omega_1$. Absorption of the signal field in SL is defined as
$A=\langle I[U(t)]\cos(\omega_1 t) \rangle_t$,
where $U(t)$ is given by (\ref{total_eq}) and averaging $\langle\ldots\rangle_t$
is performed over a period that is common for both $U_p(t)$ and $U_s(t)$. Gain corresponds to $A<0$.
\par
In our case the pump is always quasistatic ($\omega\tau<1$), but an interaction of the signal field with
miniband electrons can be both quasistatic for sub-THz frequencies or dynamic
for $\omega_1\tau\gtrsim 1$. Accordingly, to find $I(U)$ for $\omega\tau<1$ and $\omega_1\tau<1$
we can use the Esaki-Tsu eq.~(\ref{esaki-tsu_eq}), but have to use
the formal exact solution of Boltzmann transport equation \cite{chambers} if $\omega_1\tau\gtrsim 1$.
Surprisingly, in our situation both these approaches give quite similar physical results, therefore
we focus mainly on more simple quasistatic calculations and only briefly discuss a gain for
signal with $\omega_1\tau\gtrsim 1$.
\par
For a  small signal $U_1\ll U_{\omega}$, we expand $I(U)\approx
I(U_{p})+I^{\prime}(U_{p})\times U_s$ and substitute this
expression in the definition of absorption $A$. We see that
because of averaging over time the absorption $A$ is strongly
dependent on the ratio of frequencies $\omega_1$ and $\omega$.
First, if $\omega$ and $\omega_1$ are incommensurable, we have
$A=A_{inc}\equiv \langle I^{\prime}(U_{p})\rangle U_1/2$.
Obviously, for the Esaki-Tsu dependence $A_{inc}=\Lambda_p
U_1/2>0$. Moreover, $A>0$ also for the probe with
$\omega_1\tau\gtrsim 1$. Second, if $\omega_1=n\omega/2$ ($n$ is
an odd number), we found that always $A>0$ for $\Lambda_p>0$.
Thus, no gain at incommensurate or half-integer frequencies is
possible in the conditions of suppressed domains.
\par
Next, we consider the case when the probe frequency is
$\omega_1=n\omega$ (fig.~\ref{fig_device}). In this case, the
total absorption $A$ is the sum of three terms: Defined above
$A_{inc}$, as well as $A_{h}\equiv\langle I(U_{p})\cos(n\omega
t)\rangle$ and $A_{coh}\equiv\langle
I^{\prime}(U_{p})\cos(2n\omega t)\rangle U_1/2$, which can be
called the incoherent, the harmonic and  the coherent absorption
components, correspondingly \cite{cond-mat_this}.
\par
\begin{figure}
\onefigure[scale=0.5]{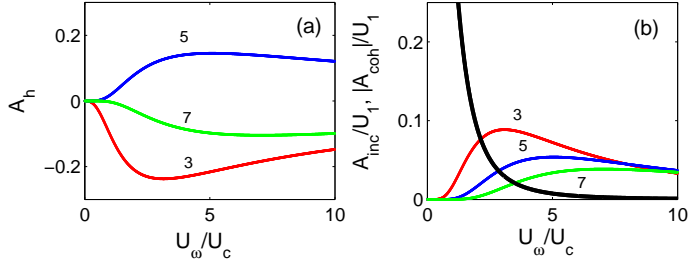}
\caption{\label{fig_harm-coh}[Color online]
(a) Amplitudes of the $3$d (red online), $5$th (blue online) and $7$th (green online)
harmonics of the current, $A_{h}$, as functions of the pump amplitude $U_\omega$.
(b) Incoherent $A_{inc}$ (black) and absolute values of coherent $-A_{coh}$ components
of small-signal absorption at $3\omega$, $5\omega$ and $7\omega$,
as functions of pump $U_\omega$.}
\end{figure}

We see that while $A_{coh}$ and $A_{inc}$ are dependent on both
$U_{\omega}$ and $U_1$, the term $A_{h}$ is a function of only
pump strength $U_{\omega}$, and therefore $A_{h}$ gives the main
contribution to the absorption of a weak probe.
For the Esaki-Tsu dependence we find
$A_{h}=(-1)^k [2( b-1)^{2k+1}]/b U_{\omega}^{2k+1}$, where
$b=(1+U_{\omega}^2)^{1/2}$ and $2k+1=n$ ($k=1,2,3\ldots$). This equation
describes just odd harmonics of the current (cf. Eq. 17 in
\cite{ignatov76}); in the limit of weak pump $U_{\omega}\ll 1$ it takes familiar form
$\propto U_{\omega}^n$ \cite{esaki71}.
Importantly, $A_{h}$ is negative for the odd values of $k$ (fig.~\ref{fig_harm-coh}a).
Therefore, generated harmonics with $n=3,7,\ldots$ can provide \textit{seeding gain} for
an amplification of a probe field. In quasistatic picture, it has no threshold in the
amplitude of pump $U_{\omega}$. Thus, if the pump amplitude
is less than the critical Esaki-Tsu voltage, the gain at $3\omega$
will be not accompanied by the space-charge instabilities in SL.
\par
Nonparametric effects in absorption are described by the term
$A_{inc}\propto\Lambda_p$. It is always positive because of
space-charge stabilization. The term $A_{coh}$ describes a
parametric amplification of the probe field due to a coherent
interaction of the pump and the probe fields in SL. Numerical
calculation of the integral demonstrates that $A_{coh}$ is always
negative for odd $n=3,5,7$; its absolute value is larger than
$A_{inc}$ for a large enough pump strengths
(fig.~\ref{fig_harm-coh}b). In the limit of weak pump
$U_{\omega}\ll 1$, we find $A_{coh}\propto U_{\omega}^{2n} U_1$.
Now we can describe the amplification of a weak signal at
$3\omega$ under the action of a weak pump. Third harmonic of a
weak pump is generated at the cubic nonlinearity of SL
characteristic, then it can get a seeding gain at the same
nonlinearity ($A_{h}\propto U_{\omega}^3$), but the next step of
parametric gain, $A_{coh}$, uses already the seventh order
($\simeq U^7$) nonlinearity in the $I(U)$ dependence.
\par
We should underline that the sign of generated current at the
frequency of harmonic plays a very important role in the process
of amplification. Really, let us compare the action of SL
oscillator tuned either to $3\omega$ or to $5\omega$ modes
(fig.~\ref{fig_device}). Net gain at $3\omega$ is quite large,
what can provide an effective operation of the oscillator.  On the
other hand, a radiation at $5$th harmonic also can be generated in
SL. However, in contrast to the case of the $3$rd harmonic, it
cannot be further amplified (because $A_{h}>0$ and
$A_{h}>|A_{coh}|$) and field in the cavity, tuned to $5\omega$,
eventually evolves to zero.
%
\section{Large-signal gain}
We calculate the large-signal absorption $A(U_{\omega},U_1)$ using
two methods: (i) simple 1D calculations with Esaki-Tsu
characteristic (eqs.~(\ref{esaki-tsu_eq}) and (\ref{total_eq}))
and (ii) 3D single-particle Monte Carlo computations with an
account of electron scattering at optical and acoustic phonons
\cite{demarina}. For the Monte Carlo computations we consider
GaAs/AlAs SL of the period $a=6.22$ nm and the miniband width
$\Delta=24.4$ meV. Static UI characteristic of this SL can be well
described by the Esaki-Tsu formula with $E_c=4.8$ kV/cm
and $\tau=220$ fs. We took $\omega/2\pi=100$ GHz. These are our
default parameters.
\par
The dependence $A(U_{\omega})$ for $3\omega$-generation and for
different values of the relative probe amplitude,
$\eta=U_1/U_{\omega}$, are shown in fig.~\ref{fig_large-signal}.
Results of simple 1D theory and 3D Monte Carlo simulations are in
a good agreement. For $\eta=0$ and for small $\eta\ll 1$, the
dependence of $A(U_{\omega})$ follows to the corresponding
dependencies for the seeding gain, $A_h(U_{\omega})$, and the
small-signal gain. With a further increase of $\eta$, the gain
decreases and finally the absorption becomes zero for some
$\eta=\eta_0$: $A(U_\omega,\eta_0)=0$. The value
$\eta_0^2=[U_1^{st}/U_\omega]^2$ determines the maximal device
efficiency for the given pump (here $U_1^{st}$ is the voltage
amplitude corresponding to the stationary amplitude of ac field in
the resonator). Inset in fig.~\ref{fig_large-signal} shows the
smooth dependence of $\eta_0$ on the pump amplitude. Generation
with low efficiency (less than 5\%) is possible even without
static NDC, for $U_{\omega}<1$. The efficiency reaches its maximal
value of $23$\% at $U_\omega\simeq 4$. For $7\omega$-oscillations
the maximum of $\eta_0^2$ is $4$\%.
\section{Numerical evidence}
\begin{figure}
\twofigures[scale=0.3]{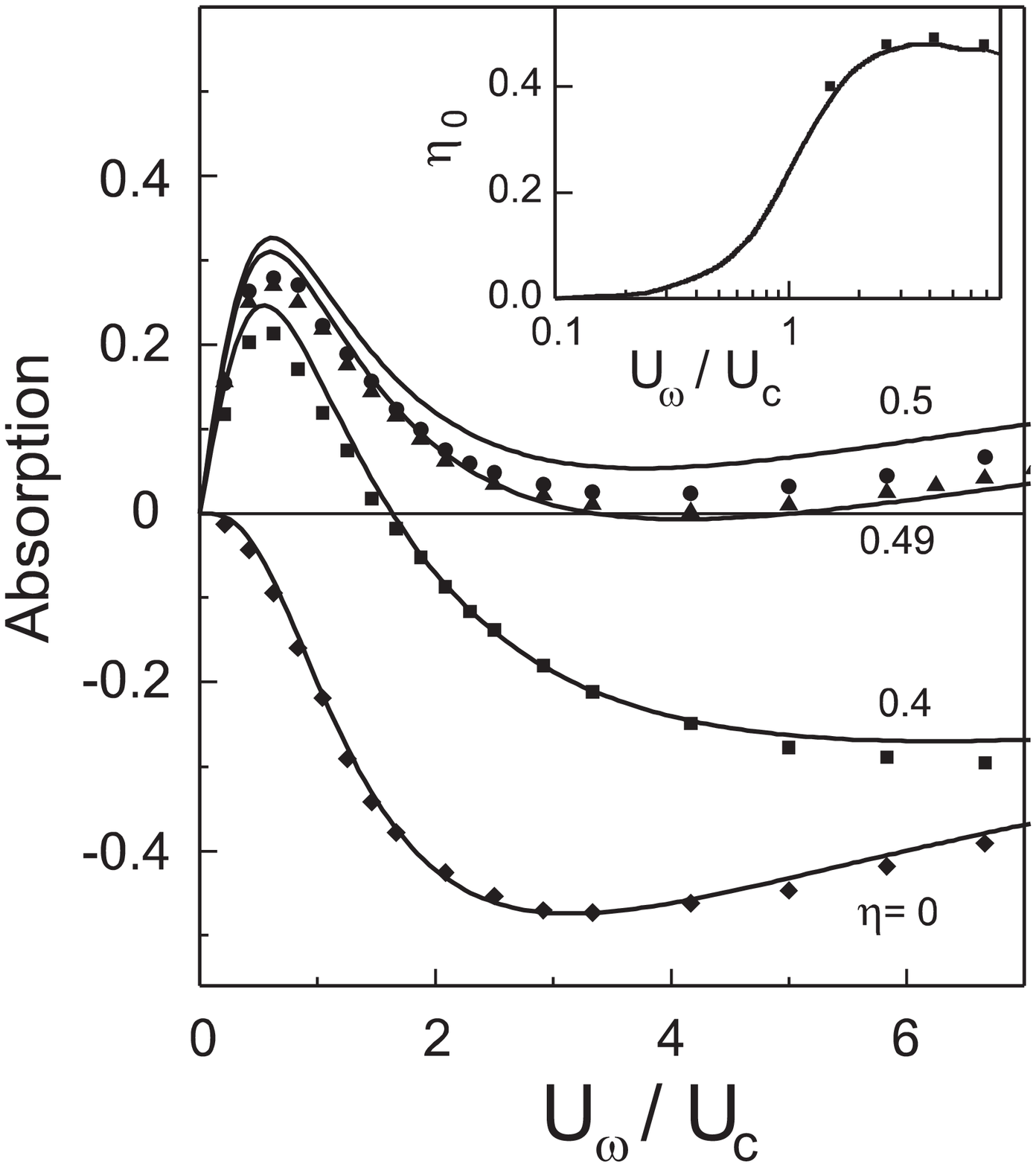}{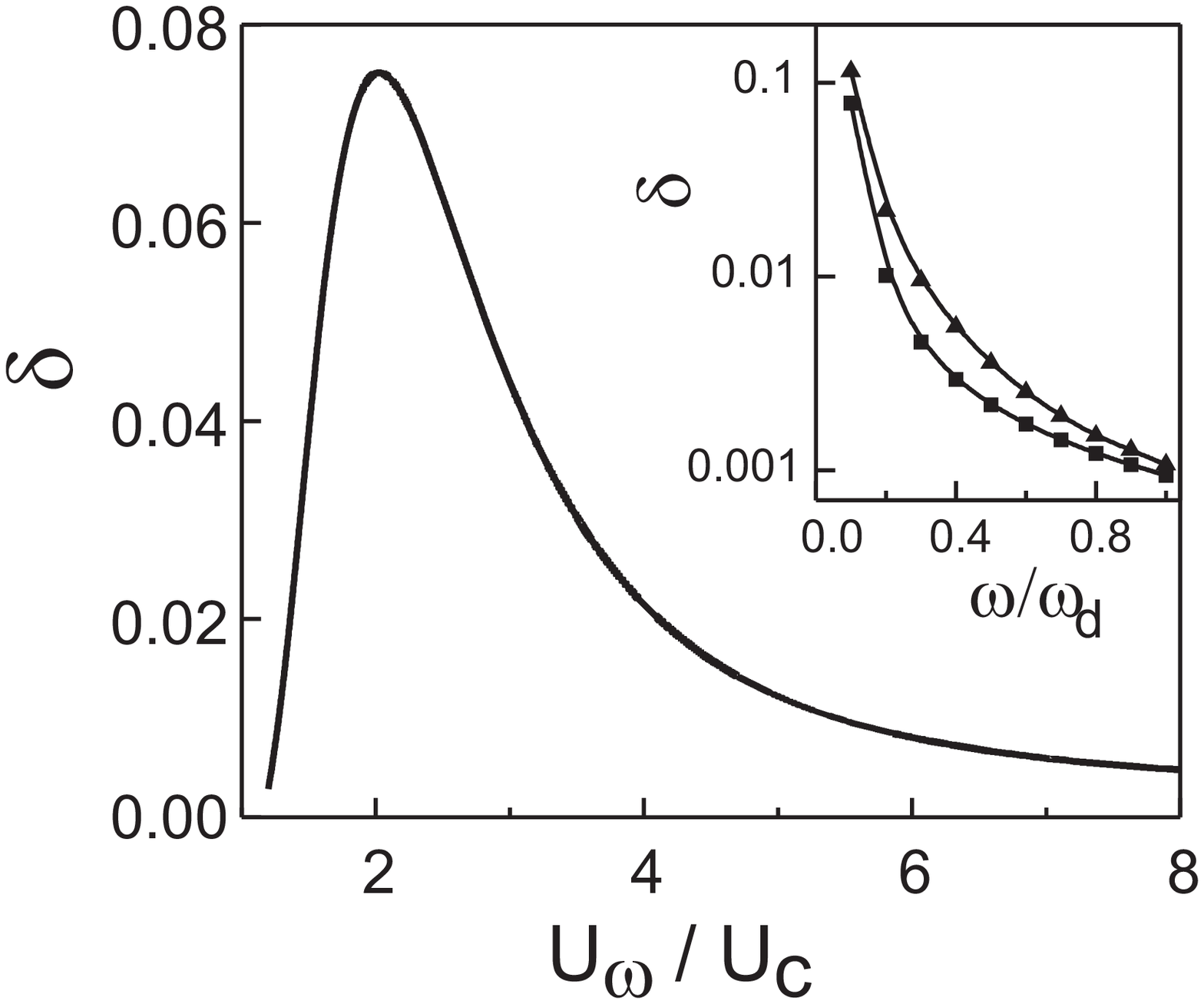}
\caption{\label{fig_large-signal} The dependence of absorption at
$3\omega$ on the pump amplitude $U_{\omega}$ for different
relative probes, $\eta$. Inset: Maximal relative amplitude of the
signal allowing gain, $\eta_0$, as a function of the pump.
Calculation with Esaki-Tsu characteristic (solid) and Monte Carlo
technique (symbols).}
\caption{\label{fig_dd-compare}
Regime of suppressed domains in SL
oscillator. Relative reduction of gain at $3$d harmonic $\delta$
vs voltage amplitude $U_\omega$ for $\omega/\omega_d=0.1$ and
$\eta=U_1/U_\omega=0.1$. Inset: $\delta$ as a function of the
ratio $\omega/\omega_d$ for $U_\omega/U_c=2$, $\eta=0$ (squares)
and $\eta=0.4$ (triangles).}
\end{figure}

We directly calculate the influence of inhomogeneous distributions
of space-charge on the gain in SL. We numerically solve the drift-diffusion model
for the time-dependent voltage $U(t)$ (eq.~(\ref{total_eq})) with
$U_{dc}=0$ and determine the absorption at the $n$th harmonic as
$A_d=\langle\langle(j(x,t)/j_0)\cos n\omega t\rangle\rangle$,
where $j_0=e V_p N$ and averaging
$\langle\langle\ldots\rangle\rangle$ is performed both over the
period $T$ and the length $L$. In the computations a spatial
fluctuation of charge distribution was caused by a local Gaussian-profiled deviation
of $E_c$ (or $\tau$). We observed a periodic spatial
accumulation of charges during a part of period $T$. We computed
the relative decrease in gain, $\delta=(A-A_d)/A$, for different
harmonics $n$ and pumps $U_\omega$. Dependence of $\delta$ on
$U_\omega$,  for SL of $130$ periods and for $\omega/\omega_d=0.1$
($\omega_d\equiv\omega_{pl}^2\tau$), is shown in
fig.~\ref{fig_dd-compare}. In fig.~\ref{fig_dd-compare} the
reduction of gain due to charge accumulation is $2$\% for optimal
$U_\omega=4$ and it is less than $8$\% overall. For our default
parameters, the value $\omega/\omega_d=0.1$ corresponds to the
doping $N=2\times 10^{16}$ cm$^{-3}$. In this situation
$\omega\tau_{dom}\simeq 1$.
We found that with an increase of $\omega/\omega_d\propto \omega\tau_{dom}$ , the value
of $\delta$ quickly decreases (see Inset in Fig.~\ref{fig_dd-compare}).
Therefore, for higher frequencies or lower doping
($\omega_d\propto N$) providing $\omega/\omega_d>0.1$, the
influence of charge accumulation on the gain is practically negligible.
\par
Additionally, we also have performed the ensemble Monte Carlo simulations \cite{ensemb_mc}
to modell the oscillator with account of space-charge dynamics in SL.
In general, these 3D simulations do confirm  our main conclusions
on the suppression of domains made in the framework of 1D drift-diffusion model.
\section{Conclusion}
In summary, we derived the necessary and sufficient conditions for
the suppression of space-charge instabilities in the
high-frequency SL oscillators with a microwave pump. Our findings
can provide simple tests to distinguish domainless and
domain-mediated regimes of generation in these oscillators, such
as a measurement of the dependence of the time-average current on
the dc voltage or a comparison of operation at $3\omega$ and
$5\omega$.
\par
The domainless oscillator can operate at THz frequencies. Probably
the most easily realizable case is to use a pump with $\omega$
near 100 GHz to obtain an output in the important frequency band
of hundreds of GHz. In order to suppress electrical domains and to
use resonator with a reasonable $Q$ (e.g., $Q\lesssim 100$ for
generation at $3\omega$), the doping of SL should be $\simeq
10^{16}$ cm$^{-3}$.
\par
In the recent experiments \cite{renk-apl,renk-prl}, a generation
at the third harmonic of a strong microwave pump of $100$ GHz have
been reported in SL device. This generation has been first
attributed \cite{renk-apl} to the process of periodic formation of
electric domains, but later it has been re-attributed to a
parametric generation in \cite{renk-prl}. In
\cite{renk-apl,renk-prl} a heavy doped ($N=10^{18}$ cm$^{-3}$) SL
was used, and therefore this device certainly operated in the
regime with $\omega\tau_{dom}\ll 1$, which is opposite to the
condition of limited charge accumulation. In this situation, a
sufficient charge accumulation during every cycle of pump field
arises, and therefore SL can switch to ANC state, as has been
shown for the case of microwave-driven Gunn diode in \cite{banis}.
We also saw such kind of ANC states in our simulations of SL
oscillator within the drift-diffusion model. This ANC induced by
the accumulated charge completely alters electric stability of SL
device and results in an operational mode that is quite different
from the mode based on an avoidance of ANC. Detailed theoretical
analysis of the experiments \cite{renk-apl,renk-prl} goes beyond
the scope of present paper.
\acknowledgments
This research was supported by Academy of Finland (grants 202697
and 109758), Emil Aaltonen Foundation, Programme of RFAE (grant
4629) and AQDJJ Programme of ESF.
\section{Erratum}
Additional modelling of a high-frequency generation with
consideration  of a resonator tuned to $n\omega$ showed that the
phase difference $\phi_n$ between the pump field ($\omega$) and
the signal field in resonator ($n\omega$) can be either $0$ for
$n=3,7$ or $\pi$ for $n=5,9$ within the quasistatic approximation.
Condition $A_h>0$ determines the phase shift $\phi_5=\pi$ between
the pump field ($\omega$) and the signal field in resonator
($5\omega$), but it does not prevent a growth of this mode, in
contrast to our previous statement. Therefore, the amplitude of
field $E_1$ in an ideal resonator tuned to $5\omega$ in fact can
grow until it reaches a stationary value $E_1^{st}$ corresponding
to zero value of the total absorption $A(E_\omega, E_1^{st})=0$,
which is now defined as $A=\langle I[E(t)]\cos(n\omega t+\phi_n)
\rangle_t$ for the total field $E(t)=E_\omega \cos(\omega
t)+E_1\cos(n\omega t+\phi_n)$. The maximal efficiency
$\eta_0^2=[E_1^{st}/E_\omega]^2$ is $10\%$. Next, following the
necessary condition for suppression of domains [eq. (3)] the
generation at $5\omega$ is still possible in domainless mode.

\end{document}